\documentclass[prc,twocolumn,showpacs,preprintnumbers,amsmath,amssymb]{revtex4}
\usepackage{graphicx}% Include figure files
\usepackage{bm}% bold math
\usepackage{booktabs}
\usepackage{subfigure}
\usepackage[usenames]{color}
\usepackage[version=3]{mhchem}
\def\gsim{\buildrel > \over {_{\sim}}}

\newcommand{\beq}{\begin{equation}}
\newcommand{\eeq}{\end{equation}}
\newcommand{\be}{\begin{eqnarray}}
\newcommand{\ee}{\end{eqnarray}}

\usepackage{sublabel}
\usepackage{epsf}
\usepackage{wrapfig}
\usepackage{epsfig}
\usepackage[usenames]{color}
\usepackage{pstricks}
\usepackage{amssymb, graphics}
\usepackage{indentfirst}
\usepackage{fancyhdr}
\usepackage[latin1]{inputenc}
\usepackage[english]{babel}
\usepackage{amssymb}  %   MATH
\usepackage{amsmath}  %   MATH
\usepackage{latexsym} %   MATH
\usepackage{amsthm}   %   MATH
\usepackage{subfigure}
\usepackage{color}

\begin{document}
%\preprint{APS/123-QED}
%%%%%%%%%%%%%%%%%%%%%%%%%%%%%%%%%%%%%%%%%%%%%%%%%%%%%%%%%%%%%%%%%%%%%%%%%
\title{Short-range correlation effects on the nuclear matrix element \\ of neutrinoless double-$\beta$ decay}
%%%%%%%%%%%%%%%%%%%%%%%%%%%%%%%%%%%%%%%%%%%%%%%%%%%%%%%%%%%%%%%%%%%%%%%%%
\author{Omar Benhar$^{1}$}
\altaffiliation[Permanent address: ]{INFN and Department of Physics, ``Sapienza'' Universit\`a di Roma, I-00185 Roma, Italy.}
\author{Riccardo Biondi$^{2,3}$}
\altaffiliation[Present address: ]{
Dipartimento di Scienze Fisiche e Chimiche, Universit\`a degli Studi dell'Aquila, I-67100  L'Aquila, Italy, 
and INFN, Laboratori Nazionali del Gran Sasso (LNGS), I-67100 Assergi (AQ), Italy.} 
%\altaffiliation[Present address: ]{Department of Physical and Chemical Sciences, Universit\`a degli Studi dell'Aquila, 
%I-00167 Coppito (AQ), Italy, and INFN LNGS, I-67010 Assergi (AQ), Italy.}
\author{Enrico Speranza$^{2,3}$}
\altaffiliation[Present address: ]{GSI, Helmholzzentrum f\"ur Schwerionenforschung, Planckstrasse 1, D-64291 Darmstadt, Germany, and 
Institut f\"ur Kernphysik, Technische Universit\"at Darmstadt, D-64289 Darmstadt, Germany.}

\affiliation
{
$^1$Center for Neutrino Physics, Virginia Polytechnic Institute and State University, 
Blacksburg, VA 24061, USA  \\
$^2$Dipartimento di Fisica, ``Sapienza'' Universit\`a di Roma, I-00185 Roma, Italy. \\
$^3$INFN, Sezione di Roma, I-00185 Roma, Italy. 
}
%%%%%%%%%%%%%%%%%%%%%%%%%%%%%%%%%%%%%%%%%%%%%%%%%%%%%%%%%%%%%%%%%%%%%%%%%
\date{\today}
%%%%%%%%%%%%%%%%%%%%%%%%%%%%%%%%%%%%%%%%%%%%%%%%%%%%%%%%%%%%%%%%%%%%%%%%%
\begin{abstract}
We report the results of a calculation of the nuclear matrix element of neutrinoless double-$\beta$ decay of $^{48}$Ca, carried out taking into account 
nucleon-nucleon correlations in both coordinate- and spin-space. 
%The inclusion of spin-spin correlations, which are known to play an important role 
%in determining the short distance structure of nuclei, leads to a mixing of the Fermi and Gamow-Teller amplitudes. 
Our numerical results, obtained using nuclear matter correlation functions, suggest that inclusion of correlations leads to 
a $\sim$ 20\% decrease of the matrix element, with respect to the shell model prediction. This conclusion is supported by the results of an independent 
calculation, in which correlation effects are taken into account using the spectroscopic factors of $^{48}$Ca obtained from an {\em ab intitio} many body 
approach.  
 
\end{abstract}
\pacs{24.40.Bw,24.40.Hc,24.10.Cn}
\maketitle
%%%%%%%%%%%%%%%%%%%%%%%%%%%%%%%%%%%%%%%%%%%%%%%%%%%%%%%%%%%%%%%%%%%%%%%%%

%%%%%%%%%%%%%%%%%%%%%%%%%%%%%%%%%%%%%%%%%%%%%%%%%%%%%%%%%%%%%%%%%%%%%%%%%
\section {Introduction}
\label{intro}

A fully quantitative approach to the calculation of the nuclear matrix element (NME) determining 
the neutrinoless double-$\beta$ ($0\nu \beta \beta$) decay rate (see, e.g., Refs. \cite{BV,closure}) requires the inclusion of correlation effects, 
not taken into account in the mean field approximation underlying the nuclear shell model.  

High-resolution electron-induced nucleon knock-out experiments have provided
unambiguous evidence of the inadequacy of the independent particle model (IPM) to describe the
full complexity of nuclear dynamics. While the peaks associated with knock-out
from shell model orbits can in fact be clearly identified in the measured missing energy spectra, the
integrated strengths, yielding the  corresponding  spectroscopic factors, 
turn out to be significantly less than the IPM predictions, independent of 
the nuclear mass number \cite{BPP,PSD}. 

Long range correlations are usually included within the framework of the Quasiparticle Random Phase Approximation (QRPA) and 
its extension, or carrying out large scale shell-model calculations. 

The effect of short range correlations is taken into account modifying the two-nucleon state obtained from the
IPM through the action of a correlation function (see, e.g., Ref. \cite{tuebingen,SRC1,ECW}).  However, existing calculations have been 
performed using somewhat oversimplified correlation functions, 
%that mainly account for the effect of the repulsive core of the NN interaction. 
depending on the magnitude of the internucleon distance only. Moreover, the results exhibit a strong dependence on the shape 
of the correlation function. According to the authors of Ref. \cite{SRC1}, different choices of the correlation functions lead to 
qualitatively different results, ranging from a $\sim$20\% suppression to a $\sim$20\% enhancement of the NME, with respect 
to the shell model result.

Nuclear matter studies carried out within 
{\em ab initio} many-body approaches, based on state-of-the-art models of the nuclear hamiltonian clearly show that the 
correlation function exhibits a complex operator structure,  
reflecting the strong spin- and isospin-dependence of the NN potential as well as its non spherically-symmetric nature.

As a first step towards the implementation of more realistic correlation functions in calculations of the $0\nu \beta \beta$ NME, 
we have studied the effect of spin- and isospin-dependent correlations on the NME of the $^{48}$Ca $\to$ $^{48}$Ti $0\nu \beta \beta$ decay
using the results of accurate nuclear matter calculations carried out within the Correlated Basis Function (CBF) approach.
The inclusion of isospin dependence allows one to take into account the differences between the correlation functions
acting in the neutron-neutron and proton-neutron channels. Moerover, spin-dependent
correlations affect the Fermi and Gamow-Teller character of the transition matrix elements, leading to a mixing of the corresponding  contributions.
Being the simplest case from the point of view of nuclear structure, the  $^{48}$Ca $\to$ $^{48}$Ti decay appears to be best suited for our exploratory 
analysis of correlation effects. In addition, searches of this decay are being carried out by the CANDLES \cite{CANDLES} 
and CARVEL \cite{CARVEL} experiments.

To gauge the robustness of our approach, we have also carried out an independent calculation, in which correlation effects have been taken into account 
through renormalisation  of the shell model states. 

In Section \ref{formalism}, after recollecting the expressions of the Fermi and Gamow-Teller contributions to 
$0\nu \beta \beta$ decay within the closure approximation, we discuss the shell-model structure of the two-nucleon matrix elements (Section \ref{matel})
and the modifications arising from the inclusion of nucleon-nucleon correlations (Section \ref{corrf}). The results of our calculations are reported in 
Section \ref{results}, while in Section \ref{summary} we summarize our findings and state the conclusions.

%Correlated Basis Function (CBF) formalism, providing the basis for the derivation of the 
%correlation function employed in this work, while the structure of the $0\nu \beta \beta$ decay NME is discussed in Section \ref{matel}. 
 
%%%%%%%%%%%%%%%%%%%%%%%%%%%%%%%%%%%%%%%%%%%%%%%%%%%%%%%%%%%%%%%%%%%%%%%%%

%%%%%%%%%%%%%%%%%%%%%%%%%%%%%%%%%%%%%%%%%%%%%%%%%%%%%%%%%%%%%%%%%%%%%%%%%
%\section{Formalism}
\section{$0\nu \beta \beta$ decay}
\label{formalism}

The half-life associated with the $0\nu \beta \beta$ decay of a nucleus of mass 
A and charge Z
\begin{equation}
({\rm A},{\rm Z}) \to ({\rm Z+2},{\rm A-2}) + 2 e^-\ ,
\end{equation}
$\tau$, can be written in the form (see, e.g., Ref. \cite{BV})
\begin{equation}
\frac{1}{\tau} = G |M|^2 \left( \frac { \langle m_{\beta\beta} \rangle  }{m_e} \right)^2 \ , 
\end{equation}
where $G$ is a phase-space factor, $m_e$ is the electron mass and the so called effective 
neutrino mass is defined in terms of neutrino mass eigenvalues and  
elements of the mixing matrix according to 
\begin{equation}
\langle m_{\beta\beta} \rangle = \left| \sum_k U^2_{ek} m_k \right|^2 \ .
\end{equation}

The NME can be cast in the form
\begin{equation}
\label{Mtot}
M = M_{{\rm GT}} - \left( \frac{g_V}{g_A} \right)^2  M_{{\rm F}} \ ,
\end{equation}
where $g_V$ and $g_A$ are the vector and axial-vector coupling constant, respectively, while 
$M_{{\rm F}}$ and $M_{{\rm GT}}$ denote the Fermi (F) and Gamow-Teller (GT) transition matrix elements.

Within the closure approximation (see, e.g., Ref.\cite{closure}) $M_{{\rm F}}$ and $M_{{\rm GT}}$ can be written 
in the general form
\begin{equation}
\label{matrixel}
 M_\alpha = \langle\Psi_f,\mathcal{J}^\pi_f|\sum_{jk}\tau^+_j \tau^+_k O^\alpha_{jk} (r)\,|\Psi_i,\mathcal{J}^\pi_i \rangle \ , 
 \end{equation}
where $\alpha =$ F, GT, $\tau^+_i$ is the charge-raising operator acting in the isospin space of the i-th nucleon and 
$\Psi_i$ and $\Psi_f$ are  the intial and final nuclear states, the total angular momentum and parity of which are labeled 
 $\mathcal{J}^\pi_i$ and $\mathcal{J}^\pi_f$.
 %, respectively. 
 
 The transition operators $O^\alpha_{jk}(r)$ are defined as
 \begin{align}
 %O^F_{jk}(r) &= \openone \, H(r) = S^F_{jk}\,H(r) \,, \\
 %O^{GT}_{jk}(r) &= (\boldsymbol{\sigma}_j \cdot \boldsymbol{\sigma}_k) \, H(r)=S^{GT}_{jk}\,H(r) \,,  
 O^F_{jk}(r) &= \openone \, H(r_{jk})  \ \ \ , \ \ \  
 O^{GT}_{jk}(r) = (\boldsymbol{\sigma}_j \cdot \boldsymbol{\sigma}_k) \ H(r_{jk})\ ,  
\end{align}
where $H(r_{jk})$ is the so-called  neutrino potential, given by 
\begin{equation}
\label{neutrino:potential}
 H(r_{jk})=R_A   \ \frac{2 }{\pi}  \int_0^{+\infty} \frac{j_0 (qr_{jk})}{q+\langle E \rangle}\  q  dq \ ,
\end{equation}
with $j_0(x) = \sin x/x$.
In the above equations, $r_{jk}=|\mathbf{r}_j - \mathbf{r}_k|$ is the magnitude of the distance between the two nucleons involved in the decay process, $R_A$ is 
the nuclear radius and $\langle E \rangle$ is the average energy of the virtual intermediate states employed in the closure approximation.

%%%%%%%%%%%%%%%%%%%%%%%%%%%%%%%%%%%%%%%%%%%%%%%%%%%%%%%%%%%%%%%%%%%%%%%%%

%%%%%%%%%%%%%%%%%%%%%%%%%%%%%%%%%%%%%%%%%%%%%%%%%%%%%%%%%%%%%%%%%%%%%%%%%%%%%
%\subsection{Matrix element of the $^{48}$Ca$\to ^{48}$Ti $0\nu \beta \beta$ decay}
\subsection{Two-body matrix elements}
\label{matel}

We assume that two neutrons of the initial state nucleus decay, while the other nucleons act as spectators. Owing to the two-body nature of the transition operators, the matrix element in 
Eq.~ \eqref{matrixel} can be reduced to a sum of products of two-body transition densities (TBTD) and antisymmetrized two-body matrix elements \cite{SRC1}
\begin{multline}
\label{tbme}
 M_\alpha = \sum_{j_1, j_2, j_1', j_2',J^\pi} TBTD\,(j_1, j_2, j_1', j_2' ;  J^\pi) \\ 
 \times \langle j'_1j'_2;\, J^\pi \ T |\, \tau^+_1 \tau^+_2 O^\alpha_{12}(r)\, | j_1 j_2;\,J^\pi \ T\rangle_a \, .
\end{multline}
%\begin{multline}
%\label{tbme}
% M_{0\nu}^\alpha = \sum_{j_1, j_2, j_1', j_2',J, \mathcal{J}^\pi} TBTD\,(j_1, j_2, j_1', j_2',J,\mathcal{J}^\pi ) \times \\
% \langle\, k'_1 l'_1 j'_1, k'_2 l'_2 j'_2; \mathcal{J}^\pi_f, T_f\,|\, \tau^+_1 \tau^+_2 O^\alpha_{12}(r) \,|\,k_1 l_1 j_1, k_2 l_2 j_2;\mathcal{J}^\pi_i, T_i \,\rangle_a \,.
%\end{multline}
Here, the indeces $1$ and $2$ label the quantum numbers of the two decaying neutrons, while $1^\prime$ and $2^\prime$  refer to the final state protons. The angular momentum of a nucleon participating in the decay is denoted $j_i$ $(i=1,2$), while $J^\pi$ and $T$ specify the total angular momentum, parity and isospin of the nucleon pair, respectively. Finally,  the notation 
$| \cdots \rangle _a$ refers to antisymmetrized two-particle states. 

The coefficients $TBTD\,(j_1, j_2, j_1', j_2' ;  {J}^\pi)$ describe how the spectator nucleons rearrange 
themselves as a result of the decay process. They are computed in a model space using an effective nucleon-nucleon interaction. 

In order to carry out the calculation, the two-body matrix element in  Eq. \eqref{tbme} must be decomposed into products of reduced matrix elements of operators acting in spin and coordinate 
space. In addition, the coordinate-space two-nucleon state is rewritten in terms of relative and center of mass coordinates,  ${\bf r}_{12}~=~{\bf r}_1 - {\bf r}_2$ and ${\bf R}_{12} = ({\bf r}_1 + {\bf r}_2)/2$, according to
\begin{align}
\label{tm}
\langle {\bf r}_1 | k_1 l_1 \rangle \langle {\bf r}_2 | k_2 l_2 \rangle &= \sum_{k,l,K,L}  \langle kl,\,KL | k_1l_1,\,k_2l_2 \rangle_\Lambda \\ 
\nonumber
& \times \langle \mathbf{R}_{12}|KL \rangle \langle \mathbf{r}_{12}|kl \rangle \ ,
\end{align}
where $k_i$ and $l_i$ are the principal and angular momentum quantum numbers, respectively,  while $\langle \ldots \rangle_\Lambda$, $\Lambda$ being the 
angular momentum of the proton pair in the final nucleus, are the coefficients 
of the Talmi-Moshinski transformation of the harmonic oscillator basis \cite{TM1,TM2}.

%%%%%%%%%%%%%%%%%%%%%%%%%%%%%%%%%%%%%%%%%%%%%%%%%%%%%%%%%%%%%%%%%%%%%%%%%

%%%%%%%%%%%%%%%%%%%%%%%%%%%%%%%%%%%%%%%%%%%%%%%%%%%%%%%%%%%%%%%%%%%%%%%%%
\subsection{Correlated wave functions}
\label{corrf}

Within CBF,  the {\em correlated} nuclear states,  $| \Psi_n \rangle$,  are obtained from the shell model eigenstates, $| \Phi_n \rangle$, through the transformation
\begin{equation}
| \Psi_n \rangle = F | \Phi_n \rangle \ ,
\end{equation}
where the operator $F$, embodying the correlation structure induced by the NN
interaction, is written in the form
\begin{equation}
\label{eq:Foperator}
F=\mathcal{S}\prod_{ij} f_{ij} \  .
\end{equation}
Note that, in general, $[f_{ij},f_{ik}] \neq 0$. As a consequence, the product
in the right hand side of Eq. \eqref{eq:Foperator} has to be  symmetrized 
through the action of the operator $\mathcal{S}$.

The two-body correlation functions $f_{ij}$,
the operator structure of which reflects the structure of the NN potential, can be cast in the 
form
\beq
f_{ij}=\sum_{m=1}^6 f^{(m)}(r_{ij}) O^{(m)}_{ij} \ ,
\label{def:corrf}
\eeq
with
\beq
O^{(m)}_{ij} = [1, (\bm{\sigma}_{i}\cdot\bm{\sigma}_{j}), S_{ij}]
\otimes[1,(\bm{\tau}_{i}\cdot\bm{\tau}_{j})]
\label{av18:2}
\eeq
where $\bm{\sigma}_{i}$ and $\bm{\tau}_{i}$ are Pauli matrices acting in
spin and isospin space, respectively,
and
\beq
S_{ij}=\frac{3}{r_{ij}^2}
(\bm{\sigma}_{i}\cdot{\bf r}_{ij}) (\bm{\sigma}_{j}\cdot{\bf r}_{ij})
 - (\bm{\sigma}_{i}\cdot\bm{\sigma}_{j}) \ .
\eeq

At lowest order of the cluster expansion scheme (see, e.g., Ref. \cite{JWC}), 
including correlations in the two body matrix element of Eq. \eqref{tbme} amounts to modifying the state describing the relative motion 
of the nucleon pair involved in the decay process, appearing in Eq. \eqref{tm}, according to 
\begin{equation}
\label{corr:1}
|kl \rangle \to f_{12}|kl \rangle \ . 
\end{equation}
Note that the above procedure can just as well be seen as a replacement of the Fermi and Gamow-Teller transition operators with the effective operators ${\widetilde O}^\alpha_{12}$,  defined as
\beq
\label{eff:op}
{\widetilde O}^\alpha_{12} = f_{12} O^\alpha_{12} f_{12} \ .
\eeq
Equation \eqref{eff:op} implies that using the correlation function defined by Eqs. \eqref{def:corrf} and \eqref{av18:2} affects the operatorial structure 
of the transition operators. To see this, consider, for example, the somewhat simplified case of a correlation function including contributions with $m \leq 4$ only. 
Because for nucleons participating in double-$\beta$ decay  $(\bm{\tau}_{1}\cdot\bm{\tau}_{2}) = 1$,  the resulting correlation functions  can be rewritten 
in the form  [see Eqs. \eqref{def:corrf} and \eqref{av18:2}]
\begin{equation}
\label{simple:f}
f_{12} = f(r_{12}) + g(r_{12}) ( \bm{\sigma}_{1}\cdot\bm{\sigma}_{2}) \ , 
\end{equation}
with
\begin{align}
\label{simple:f:2}
f (r_{12}) & = f^{(1)}(r_{12}) + f^{(2)}(r_{12})  \ , \\
\label{simple:f:3}
g(r_{12}) & = f^{(3)}(r_{12}) + f^{(4)}(r_{12}) \ .
\end{align}
From the above definitions and the relation $( \bm{\sigma}_{1}\cdot\bm{\sigma}_{2})^2 = 3 - 2 ( \bm{\sigma}_{1}\cdot\bm{\sigma}_{2})$,  
it follows that inclusion of correlations in the two-body matrix elements  
leads to the appearance of a Gamow-Teller contribution to the matrix element of $O^F_{12}$, along with a Fermi contribution to the matrix element of $O^{GT}_{12}$.
 
Substituting Eq. \eqref{simple:f}  into Eq. \eqref{eff:op} one finds
\begin{align}
{\widetilde O}^F_{12} & = [ f^2(r_{12}) + 3g^2(r_{12}) ] O^F_{12} \\ 
\nonumber
& + 2 g(r_{12}) [ f(r_{12}) - g(r_{12}) ] O^{GT}_{12} \ ,
\end{align}
and
\begin{align}
{\widetilde O}^{GT}_{12} & = [ f^2(r_{12}) - 4 f(r_{12}) g(r_{12}) + 7 g^2(r_{12}) ] O^{GT}_{12} \\  
\nonumber
& +  6 g(r_{12})[ f(r_{12}) - g(r_{12}) ] O^F_{12} \ .
\end{align}

%%%%%%%%%%%%%%%%%%%%%%%%%%%%%%%%%%%%%%%%%%%%%%%%%%%%%%%%%%%%%%%%%%%%%%%%%%%%%

%%%%%%%%%%%%%%%%%%%%%%%%%%%%%%%%%%%%%%%%%%%%%%%%%%%%%%%%%%%%%%%%%%%%%%%%%%%%%
\section{Results}
\label{results}

As stated in Section \ref{intro}, our analysisis is aimed at studying the effects of nucleon-nucleon correlations. Therefore, we have kept the complications associated with the shell model 
description of the nuclear states to a minimum. 

We focused on the reaction 
\begin{equation}
\label{reac}
 \ce{^{48}_{20}Ca} \rightarrow \ce{^{48}_{22}Ti} + 2e^-  \ ,
\end{equation}
in which both the initial and the final nucleus are in their ground states, having $\mathcal{J}^\pi=0^+$. Note that  \ce{^{48}Ca} is the lightest nucleus that can undergo double-$\beta$-decay, and its shell structure is quite simple, Z=20 and  (A-Z)=28 being both magic numbers, corresponding to closed shells. 

We consider the case in which the neutrons and protons involved in the decay process occupy the $1f_{7/2}$ shell. As a consequence, in the matrix element  
 of Eqs. \eqref{tbme} and \eqref{tm} $j_1=j_2=j_1'=j_2'=7/2$,  $k_1=k_2=k_1'=k_2'=0$ and  $l_1=l_2=l_1'=l_2'=3$.
 Numerical calculations have been carried out using the TBTD reported in Ref. \cite{brown} and harmonic oscillator wave functions corresponding to   
 $\hbar \omega = 45 {\rm A}^{-1/3} - 25 {\rm A}^{-2/3}$ MeV. The vector and axial-vector coupling constant and the average energy of Eq. \eqref{neutrino:potential} have been 
 set to the values reported in Ref.~\cite{SRC1}:  $g_V = 1$, $g_A = 1.25$ and $\langle E \rangle = 7.72$ MeV. Note that the dependence of the NME on the average energy is
quite weak. Changing the value of $\langle E \rangle$ from 2.5 MeV to 12.5 MeV results in a variation of the NME of less than 5\% \cite{SRC1}.
 
 The correlation operator employed in this work includes the components with $m \leq4$ of Eq. \eqref{def:corrf}, needed to take into account spin- and isospin-dependence.  
 The radial dependence of the functions $f^{(m)}(r_{12})$ have been obtained from a realistic nuclear hamiltonian including the Argonne $v_6^\prime$ NN potential, solving the set 
 of Euler-Lagrange equations derived from the minimization of the ground state energy of isospin-symmetric nuclear matter at equilibrium density \cite{lovato}.
 
\begin{figure}[h!]
%\begin{center}
\includegraphics[scale=0.52]{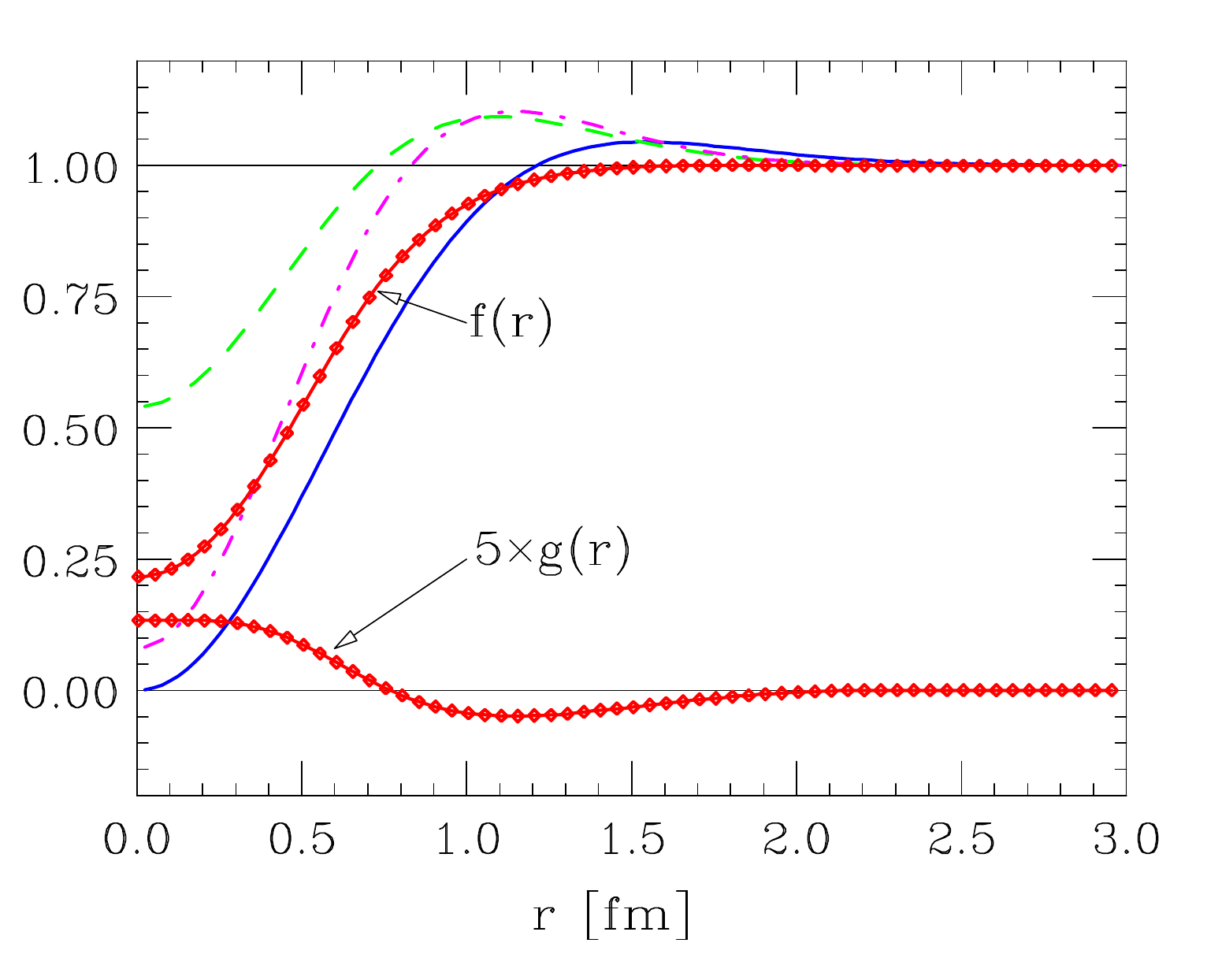}
\caption{(colour online) Radial behaviour of the correlation functions of Eq. \eqref{simple:f}. The Miller-Spencer (solid line),  AV~18 (dot-dash line) and CD~Bonn (dashed line) 
correlation functions employed in Ref. \cite{SRC1} are also shown, for comparison.}
 \label{F1}
%\end{center}
\end{figure}

 In Fig. \ref{F1} the correlation functions $f(r_{12})$ and $g(r_{12})$ (multiplied by a factor 5) of Eq.~\eqref{simple:f} are compared to those employed 
 in the study of the  $^{48}$Ca~$\to$~$^{48}$Ti $0\nu \beta \beta$ decay described in Ref. \cite{SRC1}.
 The solid, dot-dash and dashed line correspond to the correlation functions referred to as  Miller-Spencer,  AV~18 and CD~Bonn, respectively.

%%%%%%%%%%%%%%%%%%%%%%%%%%%%%%%%%%%%%%%%%%%%%%%%%%%%%%%%%%%%%%%%%%%%%%%%%%%%%
%%%%%%%%%%%%%%%%%%%%%%%%%%%%%%%%%%%%%%%%%%%%%%%%%%%%%%%%%%%%%%%%%%%%%%%%%%%%%%%%%

%%%%%%%%%%%%%%%%%%%%%%%%%%%%%%%%%%%%%%%%%%%%%%%%%%%%%%%%%%%%%%%%%%%%%%%%%%%%%%%%%
\begin{table}[h!]
\label{table}
\begin{center}
\begin{tabular}{ c c c  }
\hline
\hline
                    & $f(r_{12}) \ \ \ $   & $f(r_{12}) + g(r_{12}) (\bm{\sigma}_{1}\cdot\bm{\sigma}_{2})$ \\ 
\hline
$M /M_{SM}$ \ \ \ & 0.77 & 0.79      \\
\hline
\hline
\end{tabular}
\caption{Ratio between the $0\nu \beta \beta$ NME of Eq. \eqref{Mtot},  computed including central and central plus spin-dependent correlations
and the corresponding quantity obtained setting $f(r_{12})=1$ and $g(r_{12})=0$.}
%\label{table}
\end{center}
\end{table}
%%%%%%%%%%%%%%%%%%%%%%%%%%%%%%%%%%%%%%%%%%%%%%%%%%%%%%%%%%%%%%%%%%%%%%%%%%%%%%%%%

The numerical values of the ratio $M/M_{SM}$, where $M_{SM}$ is the NME computed without including correlations -- which amounts to setting $f(r_{ij}) = 1$ and $g(r_{ij}) = 0$  -- are listed in Table \ref{table}.
It appears that inclusion of central correlations leads to a $\gsim20$\% decrease of the NME, while the effect of spin-dependent correlations turns out to be small, and goes in the 
opposite direction.   

Our results are quite close to that obtained by the authors of Ref. \cite{SRC1} using the Miller-Spencer correlation function. 
However, the same paper also reports a $\sim$ 10\% and $\sim$ 20\% enhancement of the ratio $M/M_{SM}$, resulting from calculations 
carried out with the CD-Bonn and AV~18 correlation functions, respectively. Comparison between the shapes of the correlation functions, displayed in 
Fig.~\ref{F1} suggests that the qualitative differences in the calculated $M/M_{SM}$ ratios reflect the  differences in shape of the correlation functions. 
The enhancement of the NME, yielding $M/M_{SM}>1$,  appears to be associated with the use of correlation functions that sizeably overshoot unity at intermediate distance,
while exhibiting a less pronounced correlation hole at short distance. 

Valuable insight on the behaviour of nucleon-nucleon correlation can be obtained from theoretical studies of infinite nuclear matter. The simplifications arising from 
translation invariance allow one to carry out accurate calculations of the two-nucleon distribution functions - yielding the probability distribution of finding two
nucleons at separation distance $r$ - in both the neutron-neutron ($nn$) [or, equivalently, proton-proton ($pp$)] and proton-neutron ($pn$) channels. They are defined as
\begin{align}
\label{g:nn}
g^{nn}(r) = \frac{1}{4 \pi r^2} \langle \sum_{j>i} \delta(r-r_{ij}) \frac{1}{2} (1 - \tau^3_i) \frac{1}{2} (1 - \tau^3_j)  \rangle \  , \\
g^{pn}(r) = \frac{1}{4 \pi r^2} \langle \sum_{j>i} \delta(r-r_{ij}) \frac{1}{2} (1 + \tau^3_i) \frac{1}{2} (1 - \tau^3_j)  \rangle \  ,   
\label{g:pn}
\end{align} 
where $\tau^3_i$ is the matrix describing the $z$-component of the isospin of particle $i$, while $\langle \ldots \rangle$ denotes the ground state expectation value.

\begin{figure}[h!]
%\begin{center}
\includegraphics[scale=0.52]{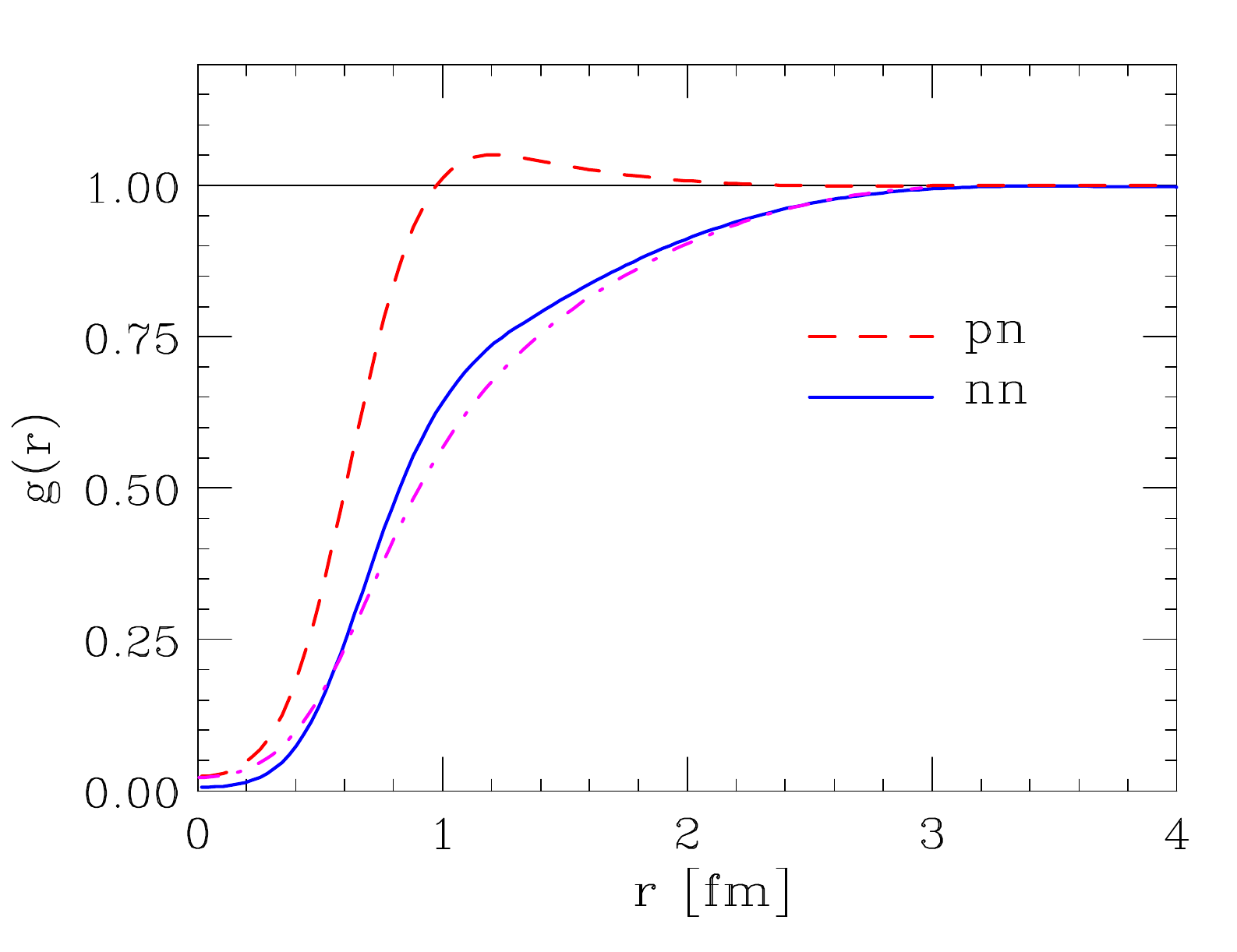}
\caption{(color online) Radial dependence of the neutron-neutron (solid line) and proton-neutron (dashed line) distribution functions of Eqs. \eqref{g:nn} and \eqref{g:pn}, computed within 
the FHNC approach using the Argonne $v_6^\prime$ NN potential \cite{lovato}. The dot-dash line corresponds to the results obtained at two-body cluster level 
using a correlation function defined as in Eqs. \eqref{simple:f}-\eqref{simple:f:3}. }
\label{F2}
%\end{center}
\end{figure}

Figure \ref{F2} shows the radial dependence of the distribution functions $g^{nn}(r)$ (solid line) and $g^{pn}(r)$ (dashed line) computed using the Fermi Hyper-Netted Chain 
(FHNC) summation scheme and the Argonne $v_6^\prime$ NN potential \cite{lovato}. The dot-dash line corresponds to the results obtained at two-body cluster level 
with the correlation function of Eqs. \eqref{simple:f}-\eqref{simple:f:3}. It clearly appears that: i) the distribution function in the $nn$ channel does not overshoot unity, and ii) 
the lowest order approximation provides a remarkably good description of the full result.

%%%%%%%%%%%%%%%%%%%%%%%%%%%%%%%%%%%%%%%%%%%%%%%%%%%%%%%%%%%%%%%%%%%%%%%%%%%%%
%%%%%%%%%%%%%%%%%%%%%%%%%%%%%%%%%%%%%%%%%%%%%%%%%%%%%%%%%%%%%%%%%%%%%%%%%%%%%%%%%

A different procedure to include correlation effects in the NME of the $0\nu\beta\beta$ decay is based on the renormalisation of the 
shell model states. Within this scheme the single nucleon state of quantum numbers $klj$ is modified according to [compare to Eq. \eqref{corr:1}]
\begin{equation}
\label{corr:2}
| klj \rangle \to \sqrt{ Z_{klj}} \  | nlj \rangle \ ,
\end{equation}
the spectroscopic factor $Z_{klj}$ being given by \cite{bisconti}
\beq
\label{def:sf}
Z^\alpha_{klj} = \int d^3x | \phi^\alpha_{klj}(x) |^2   \ ,
\eeq
where the superscript $\alpha= p,\ n$ specifies the $z$-component of the isospin, while the quasi hole wave function $\phi^\alpha_{klj}$ 
is defined as 
\begin{align}
\label{qh}
\phi^\alpha_{klj}(x_1) = \sqrt{A} \  C^\alpha_{klj} \langle \Psi^\alpha_{klj}(x_2,\ldots,x_A) | \Psi_0(x_1,\ldots,x_A) \rangle  \ .
%{\langle \Phi(x_2,\ldots,x_A | \Phi(x_2,\ldots,x_A \rangle^{1/2}
%\langle \Phi(x_2,\ldots,x_A | \Phi(x_2,\ldots,x_A \rangle^{1/2} }
\end{align}
In the above equation,  $| \Psi_0 \rangle$ and $| \Psi^\alpha_{klj}\rangle$ denote the nuclear ground state and the (A-1)-nucleon state obtained 
removing a nucleon with quantum numbers $klj$ and isospin projection $\alpha$, respectively. The normalisation factor is given by
\beq
C^\alpha _{klj}= {\langle \Psi^\alpha_{klj}| \Psi^\alpha_{klj} \rangle}^{1/2} {\langle \Psi_0 | \Psi_0 \rangle}^{1/2}  \ .
\eeq
In the absence of correlations, $Z_{klj} = 1$ for all occupied shell model states, and $Z_{klj} = 0$ otherwise.

The authors of Ref. \cite{bisconti} have carried out an {\em ab initio} calculation of the spectroscopic factors of $^{48}{\rm Ca}$ within the 
FHNC approach, using a nuclear hamiltonian including the Argonne $v_8^\prime$ NN potential supplemented with the $UIX$ 
three-nucleon  potential.

We have employed the results of Ref. \cite{bisconti} to describe correlation effects in the NME of $0\nu\beta\beta$ decay through the 
replacement
\beq
\label{sf:nme}
M_\alpha \rightarrow {\widetilde M}_\alpha = [ 1 - Z^p_{1f_{7/2}}(^{48}{\rm Ti}) ] Z^n_{1f_{7/2}}(^{48}{\rm Ca}) M_\alpha \ , 
\eeq 
which,  under the additional assumption 
\beq
\label{assumption}
[ 1 - Z^p_{1f_{7/2}}(^{48}{\rm Ti}) ] \approx  Z^n_{1f_{7/2}}(^{48}{\rm Ca}) \ , 
\eeq
yields 
\beq
\label{sf:final}
{\widetilde M}_\alpha = [ Z^n_{1f_{7/2}}(^{48}{\rm Ca}) ]^2 M_\alpha \ .
\eeq
The correspondence between the above result and the expression of the NME involving the correlation 
functions can be easily shown using correlated states in Eq. \eqref{qh}, and using the two-body cluster
approximation to evaluate the overlap.

Substitution of  the numerical value reported in Table III of Ref. \cite{bisconti} -- $Z^n_{1f_{7/2}}(^{48}{\rm Ca}) = 0.91$ -- 
in the NME of Eq. \eqref{sf:final} yields $M/M_{SM} = 0.83$, in fair agreement with the results listed in Table \ref{table}. 

Note that, although the validity of the approximation of Eq. \eqref{assumption} should be carefully investigated, nuclear matter results 
clearly support its accuracy \cite{nm:sf}. 

%%%%%%%%%%%%%%%%%%%%%%%%%%%%%%%%%%%%%%%%%%%%%%%%%%%%%%%%%%%%%%%%%%%%%%%%%%%%%

%%%%%%%%%%%%%%%%%%%%%%%%%%%%%%%%%%%%%%%%%%%%%%%%%%%%%%%%%%%%%%%%%%%%%%%%%%%%%%%%%
\section{Conclusions}
\label{summary}

We have carried out a study aimed at analysing the effects of short range NN correlations on the NME of the $0\nu\beta \beta$ decay of $^{48}$Ca. 

The results of our calculations, preformed using spin- and isospin-dependent correlation functions~--~obtained from the minimisation of the 
ground state energy of isopsin symmetric nuclear matter at equilibrium density~--~indicate that inclusion of correlations
leads to a $\sim$ 20\% descrease of the NME, with respect to the shell model prediction. 

Comparison between our results and those of Ref. \cite{SRC1} 
suggests that the radial behaviour of the correlation function plays a critical role. Using correlation functions that sizeably 
overshoot unity and exhibit a reduced  correlation hole leads to predict an enhancement, rather than a decrease, of the NME.

The approach employed to obtain the correlation functions used in our work provides a realistic description of the short range 
structure of two-nucleon states in nuclear matter, properly taking into account the differences between $nn$ and $pn$ pairs. Moreover, 
the lowest order cluster approximation appears to provide a remarkably good accuracy.

In order to gauge the robustness of our results against inclusion of finite size and shell effects on the correlation functions, we have 
also estimated the $0\nu \beta \beta$  decay NME using the spectroscopic factors of $^{48}$Ca computed in Ref. \cite{bisconti}
within the FHNC approach. 

While a more refined analysis~--~based on correlation functions obtained from the minimisation of the ground state energy of
$^{48}$Ca and including the full operator structure of Eqs. \eqref{def:corrf}-\eqref{av18:2}~--~is certainly called for,  the fair agreement between the results obtained 
from the two different approaches employed in our work suggests that the main features of correlation effects in the NME of $0\nu\beta \beta$ decay 
of $^{48}$Ca are understood at nearly quantitative level.

%%%%%%%%%%%%%%%%%%%%%%%%%%%%%%%%%%%%%%%%%%%%%%%%%%%%%%%%%%%%%%%%%%%%%%%%%%%%%

%%%%%%%%%%%%%%%%%%%%%%%%%%%%%%%%%%%%%%%%%%%%%%%%%%%%%%%%%%%%%%%%%%%%%%%%%%%%%
\section{Acknowledgements}
The authors are grateful to Alessandro Lovato for providing tables of the nuclear matter distribution functions shown in Fig. \ref{F2}, as well as for many 
illuminating discussions. The work of O.B. and R.B was supported by INFN under grant MB31.
%%%%%%%%%%%%%%%%%%%%%%%%%%%%%%%%%%%%%%%%%%%%%%%%%%%%%%%%%%%%%%%%%%%%%%%%%%%%%

%%%%%%%%%%%%%%%%%%%%%%%%%%%%%%%%%%%%%%%%%%%%%%%%%%%%%%%%%%%%%%%%%%%%%%%%%%%%%

\end{document}